\documentclass[12pt,a4paper,fleqn]{article}
\usepackage[DIV13]{typearea}
\usepackage{amsmath,amssymb,amsfonts}
\usepackage{graphicx}
\usepackage[footnotesize]{caption2}
\usepackage{cite}
\usepackage{mathrsfs}
\usepackage{bbm}
\usepackage{braket}

\newcommand{\GeV}{\:\text{GeV}}

\newcommand{\SU}{\textnormal{SU}}
\newcommand{\U}{\textnormal{U}}
\newcommand{\rep}[1]{\mathbf{#1}}
\newcommand{\Mp}{M_{\textnormal{P}}}
\newcommand{\Mg}{M_{\textnormal{GUT}}}

\newcommand{\Eqref}[1]{Eq.~\eqref{#1}}
\newcommand{\Figref}[1]{Fig.~\ref{#1}}
\newcommand{\Tabref}[1]{Tab.~\ref{#1}}
\newcommand{\Secref}[1]{Sec.~\ref{#1}}

\def\bea{\begin{eqnarray} }
\def\eea{ \end{eqnarray} }

\allowdisplaybreaks[1]

\begin{document}

\begin{titlepage}

\begin{flushright}
 IC/2010/039
\end{flushright}

\vspace*{1cm}

\begin{center}
{\Large\bf 
Supersymmetric Musings on the Predictivity of Family Symmetries \\
}

\vspace{1cm}
\renewcommand{\thefootnote}{\arabic{footnote}}

\textbf{
Kenji Kadota\footnote[1]{Email: \texttt{kadota@umich.edu}}$^{(a)}$,
J\"orn Kersten\footnote[2]{Email: \texttt{joern.kersten@desy.de}}$^{(b)}$,
and
Liliana Velasco-Sevilla\footnote[3]{Email: \texttt{lvelasco@ictp.it}}$^{(c)}$
}
\\[5mm]
\textit{\small
$^{(a)}$
Michigan Center for Theoretical Physics, University of Michigan,\\ Ann Arbor, MI, 48104, USA\\[2mm]
$^{(b)}$
University of Hamburg, II.\ Institute for Theoretical Physics,\\
Luruper Chaussee 149, 22761 Hamburg, Germany\\[2mm]
$^{(c)}$
The Abdus Salam ICTP, Strada Costiera 11, 34151 Trieste, Italy
}
\end{center}

\vspace{1cm}

\begin{abstract}
\noindent 
We discuss the predictivity of family symmetries for the soft
supersymmetry breaking parameters in the framework of supergravity. We
show that unknown details of the messenger sector and the supersymmetry
breaking hidden sector enter into the soft parameters, making it
difficult to obtain robust predictions. We find that there are specific
choices of messenger fields which can improve the predictivity for the
soft parameters.
\end{abstract}

\end{titlepage}

\newpage

% ======================================================================

\section{Introduction}

One approach towards understanding the observed fermion masses and
mixings is employing family symmetries. In a wide class of such models, one 
assigns the representations of the corresponding
group in such a way that the family symmetry forbids Yukawa couplings, 
while the matter fields couple to a number of flavon (or familon) and
vector-like messenger fields.  When these heavy messenger fields are
integrated out, one obtains an effective theory with non-renormalisable
couplings between matter and flavon fields.  Then the flavon fields develop
the vacuum expectation values (vevs) that break the family symmetry
spontaneously.  This generates non-renormalisable Yukawa couplings,
which are suppressed by a power of the small ratio of flavon vev to
messenger mass. As this power varies for different elements of the
Yukawa matrices, one can naturally obtain hierarchical fermion masses \cite{Froggatt:1978nt}.

In supersymmetric theories, family symmetries also restrict the soft
supersymmetry (SUSY) breaking parameters \cite{Abel:2001cv,Ross:2002mr},
provided that the mechanism mediating SUSY breaking to the visible
sector operates at a scale where the family symmetry is unbroken.  This
is the case for gravity-mediated SUSY breaking, for example, where the
characteristic scale is the Planck mass $\Mp$. If all matter fields transform under a three-dimensional representation
of a family symmetry, only soft scalar mass matrices proportional to the
unit matrix are allowed.
The trilinear scalar couplings have to vanish like the Yukawa couplings.

The breaking of the family symmetry leads to off-diagonal entries in the
soft mass matrices, suppressed by powers of the ratio of flavon vevs to
messenger masses.  Furthermore, non-zero trilinear couplings are
generated, which are not guaranteed to be proportional to the Yukawa
couplings.  In principle the deviations of the soft parameters from the
pattern of the Constrained Minimal Supersymmetric Standard Model (CMSSM)
can be calculated within a particular family model, and it
has been found that they can be sufficiently small to be compatible with
the experimental bounds
\cite{Ross:2002mr,Ross:2004qn,Antusch:2007re,Antusch:2008jf,Calibbi:2009ja,Calibbi:2010rf,Olive:2008vv}.
Therefore, one may expect that the SUSY flavour problem is absent even
after family symmetry breaking.
If there is a CP symmetry which is spontaneously broken together with
the family symmetry, then one can also address the SUSY CP problem \cite{Abel:2000hn,Abel:2001cv,Ross:2002mr,Ross:2004qn,Antusch:2007re,Antusch:2008jf,Calibbi:2009ja}.

Thus, in addition to explaining the fermion masses and mixings, family
symmetries could give calculable corrections to the soft SUSY breaking
parameters, which would offer additional experimental tests of family
symmetries because those soft parameters can be probed experimentally by
measuring flavour- and CP-violating observables. 

In this work, we discuss to what extent family symmetries can indeed
yield robust predictions for the soft parameters.  In \Secref{sec2},
after a general discussion of the formalism in the supergravity
framework, we argue that the predictivity is severely limited unless the
messenger sector and the SUSY-breaking hidden sector are known, as
illustrated by a concrete example in \Secref{sec:spec_example}.
Afterwards, \Secref{imppre} discusses a possibility to gain predictivity
by modifying the messenger sector, and \Secref{sec:Exp} compares the
ensuing predictions for a particular model with the experimental
constraints.

\section{Soft Parameters from Family Symmetries \label{sec2}}
\subsection{General Formalism \label{Subsec:genrem}}
Studies of how family symmetries restrict the squared masses and trilinear couplings of supersymmetric fields in the effective supergravity theory have now been brought forward for some time 
\cite{Ross:2002mr,Ross:2004qn,Antusch:2007re,Antusch:2008jf,Calibbi:2009ja,Calibbi:2010rf,Olive:2008vv}.
These studies made use of effective potentials restricted only by the symmetries of the models. 
We stress in this section that understanding the ultra-violet (UV) completion of such effective models,
in particular specifying the properties of the messenger fields, is crucial for the assessment
of the predictivity for the soft SUSY breaking parameters.   We therefore re-enumerate a specific approach which takes into account all the ingredients for the family symmetry breaking and its connection to SUSY breaking in the
supergravity context, similar in spirit to a study of Yukawa textures in \cite{Chankowski:2005jh}.

We assume that the matter fields $F$ ($F=Q,L$) and $f^c$ ($f=u,d,e,\nu$) transform under a three-dimensional irreducible representation $\rep{3}$ of a non-Abelian family symmetry. The flavons $\bar\phi$ transform under the conjugate
representation $\overline{\rep{3}}$.  Where necessary, we indicate components by the family indices $i,j = 1,2,3$
(e.g.\ $F=(Q_1,Q_2,Q_3)$ and $Q_1=(u,d)$ is the usual quark
$\SU(2)_\text{L}$ doublet of the first family).
The messengers are denoted by $\chi$ and can be either singlets or triplets of the family symmetry.

Explicitly, we take the following steps in order to
derive the observable quantities:
\begin{enumerate}
\item We start from supergravity with the superpotential
\begin{equation}
    W = W_\text{O} + W_\text{H} \;.
\end{equation}
As an example, we take
\begin{equation} \label{eq:Wfull}
    W_\text{O} =
    M_{\chi^f_0} {\bar\chi{}_0^f}_i {\chi_0^f}_i +
    M_{\chi^f_2} \bar\chi^f_2 \chi^f_2 +
    \lambda_H F_i H_f {\chi_0^f}_i + 
    \lambda_1 {\bar\chi{}_0^f}_i {\bar\phi{}_1}_i \chi^f_2 + 
    \lambda_2 \bar\chi^f_2 {\bar\phi{}_2}_i f_i^c + \mu H_u H_d \;,
\end{equation}
where $\lambda_H, \lambda_1$ and $\lambda_2$ are $\mathcal{O}(1)$
couplings,
$F, f$ and $i$ are understood to be summed over and
$H_{e,\nu} \equiv H_{d,u}$.  In this example, $\chi_0^f$ and $\chi_2^f$
transform as $\overline{\rep{3}}$ and $\rep{1}$, respectively,
under the family symmetry.  The superpotential \eqref{eq:Wfull} will
lead to some elements of the Yukawa matrices.
In order to generate the remaining elements,  realistic models contain
additional flavons and messengers with couplings and masses analogous to
the ones shown.  We will ignore this complication for the time being.

On the other hand, $W_\text{H}$ involves only hidden-sector fields 
$h_m$, viz.\ the flavons and a further field\footnote{In general, a set of fields.}
$h$ responsible for the breaking of SUSY by 
$\braket{\mathcal{F}_h} \neq 0$.
We assume that the couplings $\lambda_H$, $\lambda_1$ and $\lambda_2$
are real and do not depend on $h$.
The field $h$ has to be a singlet under the family symmetry, since
otherwise the breaking of SUSY, which involves vevs for both $h$ and
$\mathcal{F}_h$ in general, would also break the family symmetry.

The K\"ahler potential has the most general form allowed by gauge and
family symmetries,
\begin{eqnarray}
    K &=&
    \sum_\alpha C_\alpha^\dagger C_\alpha +
    \sum_{n\alpha} \frac{\zeta_{nC^\text{T}_\alpha}}{\Mp^2} \,
     |\bar\phi_n {C^\text{T}_\alpha}|^2 +
    \sum_{\alpha} \xi_{C_\alpha}(h_m) \, C_\alpha^\dagger C_\alpha
\nonumber\\
&& {}+
    \left[ Z_H(h_m) \, H_u H_d +
    \sum_n Z_{\chi_n^f}(h_m) \, \bar\chi_n^f \chi_n^f + \text{h.c.} \right]
 + {}\dots{} + K_\text{H}(h_m) \;,
\label{eq:Kmin}
\end{eqnarray}
where $C_\alpha$ is any field from the observable sector, whereas
$C^\text{T} = (F,f^c,\chi_0^{f\dagger},\bar\chi_0^f)$ refers to
family triplets only.
The functions $\xi$, $Z$ and $K_\text{H}$
are family singlets and depend only on the hidden-sector fields $h_m$.
For simplicity, we use the minimal form
$K_\text{H} = h^\dagger h + \sum_n \bar\phi_n^\dagger \bar\phi_n $ in the following, since any
different choice would not lead to qualitative changes of the
discussion.
Finally, $\zeta$ are numerical coefficients.
Family and gauge indices are not shown explicitly.
The dots represent terms that contain more than two fields $C_\alpha$ or
are suppressed by higher powers of $\Mp$.
We choose a basis where the leading-order K\"ahler potential, i.e.\ the
first term in \Eqref{eq:Kmin}, is minimal.

We assume the hierarchy of scales%
\footnote{Here and in the following, $\chi$ refers to some messenger.
 The indices identifying one particular field will be shown only if they
 are relevant.  Likewise, $\bar\phi$ will denote some flavon.}
\begin{equation} \label{eq:Scales}
    \braket{\bar\phi} \lesssim M_\chi \ll \Mp \;,
\end{equation}
because we 
\begin{enumerate}
\item need sufficiently small Yukawa couplings (with a relatively
weak hierarchy between one pair of flavon vevs and messenger masses due
to the large $y_t$), and
\item
would like to ensure that the Planck-suppressed terms with coefficients
$\zeta$ in \Eqref{eq:Kmin}, which
give rise to off-diagonal corrections to the soft masses
after the breaking of the family symmetry,%
\footnote{In contrast, the terms involving $\xi(h_m)$ do not lead to
flavour violation, since $\xi$ are singlets under the family symmetry.
}
are negligible compared to similar terms that arise after
integrating out the messengers and are suppressed by $M_\chi^2$, as we
will discuss shortly.
\end{enumerate}
\item We integrate out the messengers by solving the equations
$\partial W/\partial\chi=\partial W/\partial\bar\chi=0$
for the messenger fields \cite{Brizi:2009nn}.
In the example of \Eqref{eq:Wfull}, this yields the effective superpotential
(cf.\ \Figref{fig:Messengers33})
\begin{figure}
\centering
\includegraphics{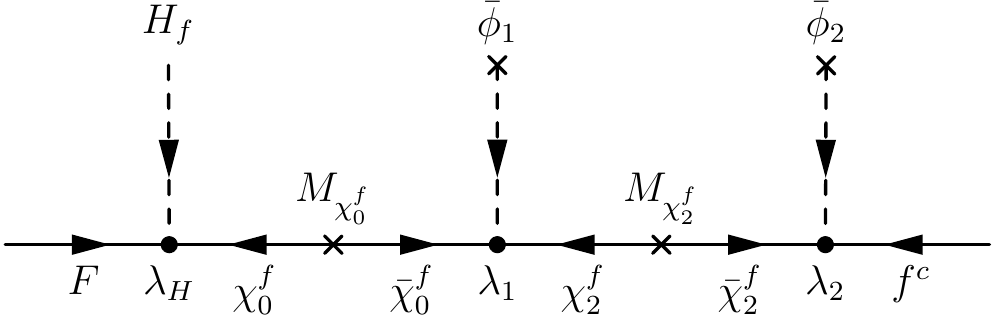}
\caption{Feynman diagram responsible for the Yukawa couplings.
 Under the SU(3) family symmetry, $\chi^f_1 \sim \overline{\rep{3}}$,
 and $\chi^f_2 \sim \rep{1}$.
}
\label{fig:Yuk_triplet_mess}
\label{fig:Messengers33}
\end{figure}
\begin{equation} \label{eq:Weff}
    W_\text{O} =
    \frac{\lambda_H \lambda_1 \lambda_2}{M_{\chi^f_0} M_{\chi^f_2}} \,
    F_j {\bar\phi{}_1}_j \, H_f \, {\bar\phi{}_2}_i f_i^c
\end{equation}
and the effective K\"ahler potential
\begin{eqnarray}
    K &=&
    F_i^\dagger F_i \left[ 1 + \xi_F +
     \frac{\lambda_H^2 \, (1+\xi_{\bar\chi_0^f})}{M_{\chi^f_0}^2} H_f^\dagger H_f
    \right]
\nonumber\\
&&{}+
    f^c_i f_j^{c\dagger} \left[ \delta_{ij} + \xi_{f^c} \delta_{ij} +
     \frac{\lambda_2^2 \, (1+\xi_{\chi_2^f})}{M_{\chi^f_2}^2}
     {\bar\phi{}_2}_i {\bar\phi{}_2^\dagger}_j
    \right] 
\nonumber\\
&&{}+
    H_f^\dagger H_f \left( 1 + \xi_{H_f} \right) +
    \left( Z_H \, H_u H_d + \text{h.c.} \right) +
    {}\dots{} + K_\text{H}
\label{eq:Keff}
\\
&\equiv&
    \tilde K_{F^\dagger_i F_j} \, F_i^\dagger F_j +
    \tilde K_{f^c_i f^{c\dagger}_j} \, f^c_i f^{c\dagger}_j +
    \tilde K_{H_f^\dagger H_f} \, H_f^\dagger H_f + 
    \left( Z_H \, H_u H_d + \text{h.c.} \right) +
    {}\dots{} + K_\text{H} \;,
\nonumber
\end{eqnarray}
where we have omitted $\Mp$-suppressed terms proportional to $\zeta$ and
terms suppressed by higher powers of messenger masses.
Thus, we obtain non-minimal terms suppressed by messenger masses.
This can be visualised by diagrams like the one shown in
\Figref{fig:Kaehler33}.
\begin{figure}
\centering
\includegraphics{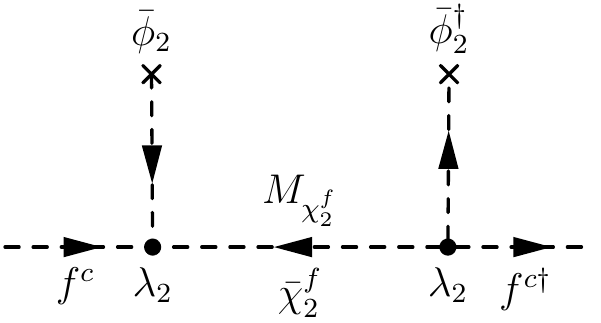}
\caption{Diagram relevant for the K\"ahler potential.}
\label{fig:Kaehler33}
\end{figure}
Note that in this way the messengers are integrated out already around
$\Mp$, not only at their own mass scales.  This should not be a problem
as long as we do not aspire high-precision calculations including the
running of parameters between $\Mp$ and $M_\chi$.

\item From the effective potentials we calculate the scalar potential.
It contains
$\frac{\partial W_\text{O}}{\partial \bar\phi}$, which yields an
important contribution to the trilinear scalar couplings. The minimisation
of the potential yields vevs for all hidden sector fields and their
$\mathcal{F}$ terms, breaking both SUSY and the family symmetry.

\item We take the flat limit, i.e.\ $\Mp\to\infty$ and 
$m^2_{3/2}=\braket{e^{K_\text{H}/\Mp^2} |W_\text{H}|^2}/\Mp^4=\text{const.}$ \cite{Cerdeno:1998hs}. 
This removes the dynamical degree of freedom $h$ from the theory.  In
contrast, both the flavon vevs $\braket{\bar\phi}$ and the dynamical
fields $\bar\phi$ are still present, since they have couplings to the observable sector that are suppressed by $M_\chi$ rather than $\Mp$. It is only at the scale $\braket{\bar\phi}<\Mp$ that they decouple. Again, this should not be a problem as long as we do not aim to calculate the running of parameters between $\Mp$ and $\braket{\bar\phi}$.

\item We rescale the superpotential of the visible sector,
\begin{equation} \label{eq:DefN}
    W'_\text{O} =
    W_\text{O} \left\langle
     \frac{W_\text{H}^*}{|W_\text{H}|} \, e^{\frac{1}{2\Mp^2} \sum_m|h_m|^2}
    \right\rangle \equiv
    \mathcal{N} \, W_\text{O} \;.
\end{equation}
This is necessary in order to obtain the usual globally supersymmetric
contribution
$\sum_\alpha |\partial W'_\text{O} / \partial C_\alpha|^2$
to the scalar potential.
The rescaling is absorbed in the effective Yukawa couplings,
\begin{equation} \label{eq:YHat}
Y'_{f^c_i F_j H_f} \equiv
    \mathcal{N} \, Y_{f^c_i F_j H_f} \equiv
    \mathcal{N} \, \lambda_H \lambda_1 \lambda_2 \,
    \frac{\braket{\bar\phi_2}_i \braket{\bar\phi_1}_j }{ M_{\chi^f_0} M_{\chi^f_2} }
\;.
\end{equation}
$Y_{f^c_i F_j H_f}$ denotes the $ij$ component of the matrix $Y_{\alpha\beta\gamma}$ coupling the fields $C_\alpha=f^c$, $C_\beta=F$ and $C_\gamma=H_f$. Note that the rescaled Yukawa couplings $Y'$ are the ones directly related to observable quantities (up to canonical normalisation) that are determined by the
fit to the fermion masses.

\item
The scalar potential now consists of the globally supersymmetric
part and soft SUSY breaking terms.  Assuming that no $\mathcal{D}$ terms
contribute to SUSY breaking, we determine the latter using
Eqs.~(11,\,12) of \cite{Brignole:1997dp}, which in our notation become
\begin{subequations} \label{eq:Brignole}
\begin{eqnarray}
    m^{\prime2}_{\bar{\alpha}{\beta}} &=&
    m_{3/2}^2 \braket{\tilde K_{\bar{\alpha}{\beta}}} -
    \Braket{\mathcal{F}^{*\bar{m}} \left(   \partial^*_{\bar{m}}\partial_n  \tilde K_{\bar{{\alpha}}{\beta}}- (\partial^*_{\bar{m}} \tilde{K}_{\bar{{\alpha}}{\gamma}}) \,\tilde{K}^{\gamma\bar\delta} \,\partial_n \tilde{K}_{\bar{\delta}{\beta}} \right){\mathcal{F}}^n } ,
\\
    a'_{{\alpha}{\beta}{\gamma}} &=&  \braket{
    {\mathcal{F}}^m } \left[\Braket{
     \frac{\partial_m K_\text{H}}{\Mp^2}} Y'_{{\alpha}{\beta}{\gamma}} +
     \frac{{\mathcal{N}}\partial  Y_{{\alpha}{\beta}{\gamma}}}{\partial {\braket{h_m}}} \right] \nonumber\\
&&{}-  \braket{
    {\mathcal{F}}^m }\left[
\Braket{\tilde{K}^{\delta\bar\rho} \, (\partial_m \tilde K_{\bar\rho{\alpha}})} Y'_{\delta{\beta}{\gamma}}+ ({\alpha} \leftrightarrow {\beta})+ ({\alpha} \leftrightarrow {\gamma})    
    \right] ,
\label{eq:BrignoleTrilinears}
\end{eqnarray}
\end{subequations}
where
$\tilde K_{\bar\alpha\beta} \equiv \frac{\partial^2 K}{\partial C_{\bar\alpha}^\dagger \partial C_\beta}$
with $C = (F, f^{c\dagger}, H_f)$
and where $\tilde{K}^{\gamma\bar\delta}$ denotes the elements of the
inverse matrix.
Besides,
$\partial_m\equiv\partial/\partial h_m$,
$\partial^*_{\bar m}\equiv\partial/\partial h_{\bar m}^*$, and e.g.\
$\braket{\mathcal{F}^{\bar\phi_1}} \partial/\partial\bar\phi_1 \equiv
 \braket{\mathcal{F}^{{\bar\phi{}_1}_i}} \partial/\partial{\bar\phi{}_1}_i$.
We have expressed the formula for the trilinear couplings in terms
of $Y'$ for convenience, where it is possible without ambiguity.
Primes denote parameters before canonical normalisation.
There are different $\mathcal{F}$-term vevs associated to each flavon,
$\braket{\mathcal{F}^{\bar\phi_n}}=c_n m_{3/2} \braket{\bar\phi_n}$
\cite{Abel:2001cv,Ross:2002mr}, where $c_n\neq c_m$ for $n\neq m$.%
\footnote{
Here we use
$\braket{\mathcal{F}^m}=\braket{e^{K/(2\Mp^2)}\frac{|W_\text{H}|}{\Mp^2}}\braket{ K_\text{H}^{m\bar n}  (K_{\bar n} + \frac{W^*_{\bar n} }{W^*} )}$. 
For the flavons $\braket{|{\mathcal{F}}^{\bar\phi_n}|^2}$ behaves as $m^2_{3/2} c^2_n \left|\braket{K_{\bar\phi_n} +\frac{W_{\bar\phi_n}}{W}}\right|^2$, then it is assumed that the term containing $|K_{\bar\phi_n}|^2$ is the dominant one. Formally the coefficients $c_n$ should be determined from the process that sets completely the minimum
of the scalar potential and so depends on details of how SUSY is broken. However, since the $\mathcal{F}$ terms in general are proportional to $\bar\phi_n$ the coefficients $c_n$ are expected to be ${\mathcal{O}}(1)$.
}

As mentioned, we are treating the flavons as part of the hidden sector associated to the breaking of SUSY and therefore there are also non-zero vevs for their $\mathcal{F}$ terms, although they are not the main contribution to SUSY breaking, the leading source of course being the family-blind field $h$. It is also important to note that if there was only one flavon in the theory and thus only one $\mathcal{F}$ term, then we can immediately see from Eqs.~\eqref{eq:Brignole} that when going to the canonical basis there would be no off-diagonal terms, even with a non-trivial K\"ahler metric.  On the other hand it can be quickly computed \cite{Ross:2004qn} that with at least two different flavons and consequently different $\mathcal{F}$ terms, the soft mass matrices have the same structure as the K\"ahler metric but with different $\mathcal{O}(1)$ coefficients in each component,
\begin{equation} \label{eq:EstimateSoftMasses}
    m^2_{\tilde f^c_i \tilde f_j^{c\dagger}} \sim
    \mathcal{O}(1) \, m_{3/2}^2 \braket{\tilde K_{f^c_i f^{c\dagger}_j}} ,
\end{equation}
where of course the precise values of the $\mathcal{O}(1)$ coefficients depend on the details of the K\"ahler potential and the $\mathcal{F}$ terms.

\item We normalise the visible-sector fields to obtain
canonical kinetic terms,
\begin{equation}
    F \to \hat F \equiv V_F^{-1} F
    \quad,\quad
    f^c \to \hat f^c \equiv f^c \, {V_{f^c}^{-1}}^\dagger
    \quad,\quad
    H_f \to \hat H_f \equiv \tilde K_{H_f^\dagger H_f}^\frac{1}{2} \, H_f \;,
\end{equation}
where the (non-unitary) matrices $V$ diagonalise the K\"ahler metric,%
\footnote{At the order we are considering the K\"ahler potential does
not mix different fields $F$ or $f^c$.  Hence, every block
$\tilde K_{F^\dagger F}$ and $\tilde K_{f^c f^{c\dagger}}$ in the K\"ahler metric
can be diagonalised with a different matrix.
Likewise, the block associated to the Higgs fields is diagonal.
We use $\tilde K_{F^\dagger F}$ to denote the matrix whose $ij$ element is
$\tilde K_{F_i^\dagger F_j}$, and analogously for other quantities.}
\begin{eqnarray}
V_F^\dagger \tilde K_{F^\dagger F} V_F = \mathbbm{1} \;,
\quad
V_{f^c}^\dagger \tilde K_{f^c f^{c\dagger}} V_{f^c} = \mathbbm{1} \;.
\label{eq:VfcK}
\end{eqnarray}
Consequently, the transformations of the soft parameters and the Yukawa couplings are given by 
\begin{subequations}\label{eq:CanonicalNorm}
\begin{eqnarray}
    m^{\prime2}_{\tilde F^\dagger \tilde F} &\to&
    \hat m^2_{\tilde F^\dagger \tilde F} \equiv
    V_F^\dagger \, m^{\prime2}_{\tilde F^\dagger \tilde F} \, V_F \;,
\label{eq:CanonicalNormSoftMasses}
\\
    m^{\prime2}_{\tilde f^c \tilde f^{c\dagger}} &\to&
    \hat m^2_{\tilde f^c \tilde f^{c\dagger}} \equiv
    V_{f^c}^\dagger \, m^{\prime2}_{\tilde f^c \tilde f^{c\dagger}} \, V_{f^c}
    \;,
\\
    a'_{\tilde f^c \tilde F H_f} &\to&
    \hat a_{\tilde f^c \tilde F H_f} \equiv
    \tilde K_{H_f^\dagger H_f}^{-\frac{1}{2}}
    V_{f^c}^\dagger \, a'_{\tilde f^c \tilde F H_f} \, V_F \;,
\label{eq:transftril}
\\
     Y'_{f^c F H_f} &\to&
     \hat Y_{f^c F H_f} \equiv
    \tilde K_{H_f^\dagger H_f}^{-\frac{1}{2}}
    V_{f^c}^\dagger \, Y'_{f^c F H_f} \, V_F \;.
\label{eq:transf_yuk_cb}
\end{eqnarray}
\end{subequations}

\item 
Flavour-violating parameters are computed in the super-CKM (SCKM) basis where the Yukawa couplings are diagonal,
\begin{equation}
    \widetilde Y_{f^c F H_f} =
    {U_\text{R}^f}^\dagger \hat Y_{f^c F H_f} U_\text{L}^f =
    \text{diag} \;,
\label{eq:transf_yuk_diag}
\end{equation}
and we have the corresponding transformations for the soft terms,
\begin{subequations}
\begin{eqnarray}
    \widetilde a_{\tilde f^c \tilde F H_f} &=& {U_\text{R}^f}^\dagger \hat a_{\tilde f^c \tilde F H_f} U_\text{L}^f \;,\\
    \widetilde m^2_{\tilde f,\text{LL}} &=& {U_\text{L}^f}^\dagger \hat m^2_{\tilde F^\dagger \tilde F} U_\text{L}^f \;,\\
    \widetilde m^2_{\tilde f,\text{RR}} &=& {U_\text{R}^f}^\dagger \hat m^2_{\tilde f^c \tilde f^{c\dagger}} U_\text{R}^f \;.
\end{eqnarray}
\end{subequations}

\end{enumerate}

In summary, we would like to emphasise two crucial points for the predictivity of these scenarios.
A first consequence of the supergravity formalism, including a UV completion with both a sector breaking SUSY and a sector breaking the family symmetry, is the explicit form \eqref{eq:YHat} of the Yukawa couplings, containing information on both sectors.  In the supergravity literature the dependence on the family-blind sector is a well-known fact.  However, so far this has not been considered in works studying family symmetries in the effective theory approach.  Second, the relations \eqref{eq:Brignole} between the parameters describing the Yukawa couplings and those responsible for the soft parameters are sensitive to many details of the UV completion, as we shall illustrate in the following sections.

\subsection{A Conventional Example with a Triplet Messenger\label{sec:spec_example}}
Let us consider the model presented in \cite{deMedeirosVarzielas:2005ax}
with an $\SU(3)\times\U(1)\times\U(1)'$ family symmetry as a conventional 
example.  Besides the fields mentioned in the previous
section, another flavon $\bar\phi_3$ and additional family-singlet
messengers $\chi_1^f$, $\chi_3^f$ are present.  The superpotential is a
straightforward generalisation of \Eqref{eq:Wfull}.%
\footnote{We do not mention several other fields and many details which
 are important for the model but not relevant for our discussion.
 For example, an extra Higgs field $H_{45}$, whose vev is proportional
 to the hypercharge, plays a role in generating the $23$-blocks of the
 Yukawa matrices.
}
The Yukawa couplings stem from diagrams of the type shown in
\Figref{fig:Messengers33} \cite{Varzielas:2008kr}, involving 
$\chi_0^f$ and one more messenger $\chi_n^f$ with $n=1,2,3$.
The flavons develop vevs 
$\braket{\bar\phi_3} \propto (0,0,1)$,
$\braket{\bar\phi_2} \propto (0,1,-1)$, and
$\braket{\bar\phi_1} \propto (1,1,1)$.
We assume the hierarchy
\begin{equation} \label{eq:condit_flav_hier}
\frac{(\braket{\bar\phi_3}_3)^2}{M_{\chi_3^f}} \gg
\frac{(\braket{\bar\phi_2}_2)^2}{M_{\chi_2^f}} \gg
\frac{\braket{\bar\phi_1}_2\braket{\bar\phi_2}_2}{M_{\chi_1^f}} \sim
\frac{\braket{\bar\phi_1}_2\braket{\bar\phi_2}_2}{M_{\chi_2^f}} \;.
\end{equation}
Then the Yukawa matrices are approximately given by
{\renewcommand{\arraystretch}{1.5}
\begin{equation}
Y^\prime_{f^c F H_f} \approx \mathcal{N} \lambda_H
\left(
\begin{array}{ccc}
0 & 
\lambda_1 \lambda_2 \frac{\braket{\bar\phi_1}_1\braket{\bar\phi_2}_2}{M_{\chi^f_0} M_{\chi^f_1}} &
\lambda_1 \lambda_2 \frac{\braket{\bar\phi_1}_1\braket{\bar\phi_2}_3}{M_{\chi^f_0} M_{\chi^f_1}}
\\
\lambda_1 \lambda_2 \frac{\braket{\bar\phi_2}_2\braket{\bar\phi_1}_1}{M_{\chi^f_0} M_{\chi^f_2}} &
\lambda_2^2 \frac{(\braket{\bar\phi_2}_2)^2}{M_{\chi^f_0} M_{\chi^f_2}} &
\lambda_2^2 \frac{\braket{\bar\phi_2}_2\braket{\bar\phi_2}_3}{ M_{\chi^f_0} M_{\chi^f_2}}
\\
\lambda_1 \lambda_2 \frac{\braket{\bar\phi_2}_3\braket{\bar\phi_1}_1}{M_{\chi^f_0} M_{\chi^f_2}} &
\lambda_2^2 \frac{\braket{\bar\phi_2}_3\braket{\bar\phi_2}_2}{ M_{\chi^f_0} M_{\chi^f_2}} &
\lambda_3^2 \frac{(\braket{\bar\phi_3}_3)^2}{M_{\chi^f_0}M_{\chi^f_3}} 
\end{array}
\right) ,
\label{eq:effec_yuk_r_examp}
\end{equation}
}%
which can fit the appropriate fermion masses and mixings \cite{Roberts:2001zy,Calibbi:2010rf}.
  For brevity, we use a single value
$\lambda_n$ for all the $\mathcal{O}(1)$ couplings of each flavon
$\bar\phi_n$, and likewise a single $\lambda_H$.  In general, one could distinguish between many different
values, introducing $\lambda_n^{f0m}$ for the coupling between
$\bar\phi_n$, $\bar\chi_0^f$ and $\chi_m^f$, and $\lambda_n^{fm}$ for
the coupling between $\bar\phi_n$, $\bar\chi_m^f$ and $f^c$.  However,
precisely keeping track of these complications is not necessary, since
all predictions will only be up to $\mathcal{O}(1)$ factors.

For example, up to factors of order unity one obtains
\begin{equation}
    Y^\prime_{d^c Q H_d} \sim
        \begin{pmatrix}
        0 & \epsilon_d^3 & -\epsilon_d^3 \vphantom{\epsilon^3_{d_1}}\\
        \epsilon_d^3 & \epsilon_d^2 & -\epsilon_d^2 \\
        -\epsilon_d^3 & -\epsilon_d^2 & 1 \\
        \end{pmatrix}
\label{eq:yuk_r_examp_eps}
\end{equation}
for the down-type quarks, which is compatible with observations for
$\epsilon_d \sim 0.13$ \cite{Calibbi:2010rf}.

We see that each
Yukawa coupling depends on a product of two different masses,
for instance
\begin{equation} \label{eq:Y23}
    Y'_{f^c_2 F_3 H_f} =
    \mathcal{N} \lambda_H \lambda_2^2 \,
    \frac{\braket{\bar\phi_2}_2\braket{\bar\phi_2}_3}{M_{\chi^f_0} M_{\chi^f_2}}
    \equiv -\epsilon_f^2 \;.
\end{equation}
On the other hand, due to the quantum numbers
under the family symmetry, the family triplet messengers
$\chi^f_0$ cannot occur in the diagrams relevant for the K\"ahler
potential.  Therefore, the family-dependent terms in $K$
depend only on one messenger mass,%
\footnote{The second messenger mass does appear in the term involving
 $F$ and $H$, but this term is family-blind and its contribution to the SUSY breaking parameters is
 suppressed by $\braket{H_f}^2 M_{\chi_0^f}^{-2}$.  Besides, there is a
 term proportional to
 $|\bar\phi_1|^2 |\bar\phi_2|^2 \, M_{\chi^f_0}^{-2} M_{\chi^f_2}^{-2}$,
 which is also too small to play a significant role.
} for example, from \Eqref{eq:Keff}, 
\begin{equation} \label{eq:EpsilonTilde}
    \tilde K_{f^c_2 f^{c\dagger}_3} =
    \lambda_2^2 \, (1+\xi_{\chi_2^f}) \,
    \frac{\braket{\bar\phi_2}_2\braket{\bar\phi_2^\dagger}_3}{M_{\chi^f_2}^2}
    \equiv
    -\tilde\epsilon_f^2 \;.
\end{equation}
More generally, the effective superpotential operators responsible for
Yukawa couplings, which can be determined by a fit to the fermion
masses, depend on a different combination of messenger masses than the
effective operators in the K\"ahler potential.  Consequently, the
couplings of the latter operators are \emph{not} determined.
This has profound implications for the predictivity of the theory for
the SUSY breaking parameters.

In order to study this issue without introducing unnecessary
complications, let us consider only those contributions to the soft
parameters which stem from the flavons $\bar\phi_1$ and $\bar\phi_2$.  The calculation
is performed as specified in the previous subsection.  Using the
effective K\"ahler potential \eqref{eq:Keff} and the Yukawa couplings
\eqref{eq:effec_yuk_r_examp}, we obtain contributions to some
elements of the soft masses and trilinears.  Further elements are easily
obtained by exchanging $\bar\phi_1$ and $\bar\phi_2$ or by replacing
$\bar\phi_1$ with $\bar\phi_2$ (and correspondingly for the associated
parameters $\lambda_n$, $c_n$ and $M_{\chi_n^f}$).

Let us first define two limiting cases (where ``fixed'' means fixed to
yield the correct Yukawa couplings).
\begin{description}
\item[Case 1] $\quad M_{\chi^f_0} M_{\chi^f_2} = \text{fixed}, \;
 M_{\chi^f_0} \gtrsim \braket{\bar\phi_2} \quad \Rightarrow \quad 
 M_{\chi^f_2} \gg \braket{\bar\phi_2} \;,\; \tilde\epsilon_f \ll 1$\\
The messenger mass $M_{\chi^f_0}$ is not much larger than the flavon
vev.  Then the second messenger mass $M_{\chi^f_2}$ has to be very
large, and $K$ is family-blind to a good approximation.

\item[Case 2] $\quad M_{\chi^f_0} M_{\chi^f_2} = \text{fixed}, \;
 M_{\chi^f_2} \gtrsim \braket{\bar\phi_2} \quad \Rightarrow \quad 
 M_{\chi^f_0} \gg \braket{\bar\phi_2} \;,\; \tilde\epsilon_f \lesssim 1$\\
As $M_{\chi^f_2}$ is now rather small relative to the vev,%
\footnote{In order to reach $\tilde\epsilon_f \sim 1$, \Eqref{eq:Scales}
has to be relaxed a bit, allowing for $M_{\chi_2^f} < \braket{\bar\phi_3}$.
This seems reasonable, since $\chi_2^f$ does not couple to $\bar\phi_3$,
so that no dangerous terms containing $\braket{\bar\phi_3}/M_{\chi_2^f}$
(without any accompanying suppression factors) can arise.  If instead
$M_{\chi_2^f} \ge \braket{\bar\phi_3}$, one finds that
$\tilde\epsilon_f$ cannot become much larger than $\epsilon_f$.}
we find
significant deviations from a family-blind K\"ahler potential.
\end{description}
The setup under consideration allows us to choose both of the above
cases, so that it does not predict which one is realised.  The question
is then how much the soft parameters differ between these cases.

The soft scalar masses depend only on the $\mathcal{F}$ terms and on
the K\"ahler potential
\cite{Brignole:1997dp}, which is very different in our two limiting
cases.  Canonical normalisation does not lead to any qualitative change
here, since the corresponding transformations
\eqref{eq:CanonicalNormSoftMasses} are determined by $\tilde K$ only and
not by parameters involving the Yukawa couplings. 
Therefore, we have to conclude that in fact the unknown parameters in
the K\"ahler potential prevent us from predicting the soft scalar
masses.

More precisely, we cannot predict the soft masses of
the superpartners of the right-handed fermions.  We do find that there
are no corrections to the $\SU(2)_\text{L}$ doublet scalar mass matrix
$m^2_{\tilde F^\dagger \tilde F}$, since there are no $\SU(2)_\text{L}$ doublet
messengers \cite{Antusch:2007vw}.%
\footnote{A purely left-handed (i.e.\ $\SU(2)_\text{L}$ doublet)
 messenger sector is excluded because in this case the $\SU(2)_\text{L}$
 symmetry would lead to $\epsilon_u=\epsilon_d$.
}

The structure of the trilinear scalar couplings before canonical
normalisation is
\begin{subequations}
\begin{eqnarray}
a'_{\tilde f^c_i \tilde F_j H_f} &=& m_{3/2}\left[T_h\frac{|\braket{h}|^2}{\Mp^2} + T_{\tilde f^c_i \tilde F_j H_f}   \right] Y'_{f^c_i F_j H_f}
\nonumber\\ 
&& {}-  m_{3/2} c_2 \lambda_2^2 \frac{\braket{\bar\phi_2}_i}{M_{\chi^f_2}^2} \, \sum_k \braket{\bar\phi_2^\dagger}_k \left(1-{\mathcal{O}}\left(\tfrac{|\braket{h}|^2}{\Mp^2} \right) \right) Y'_{f^c_k F_j H_f} \;,
\label{eq:tri_coups_nonpr}
\end{eqnarray}
where
\begin{eqnarray}
T_{h} &=& 1 + \mathcal{O}(\lambda_H) + \mathcal{O}(\lambda_1) + \mathcal{O}(\lambda_2) \;,
\\
T_{\tilde f^c_i \tilde F_j H_f} &=& c_1 \, p_1^{ij} + c_2 \, p_2^{ij} \;,
\label{eq:def_par_trilinears}
\end{eqnarray}
\end{subequations}
and where we have neglected terms suppressed by
$\braket{\bar\phi}^2/\,\Mp^2$.  Besides, $p_n^{ij}$ stands for the
power of $\braket{\bar\phi_n}$ that appears in $Y'_{f^c_i F_j H_f}$,
e.g.\ $p_2^{23}=2$.
The first term in \Eqref{eq:tri_coups_nonpr},
from $\partial_h K_\text{H}$ and the second line in
\Eqref{eq:BrignoleTrilinears}, gives a family-blind contribution
proportional to $Y'_{f^c_i F_j H_f}$.  We define the familiar CMSSM-like
parameter 
\begin{equation}
 A_0 \equiv m_{3/2} T_h \frac{|\braket{h}|^2}{\Mp^2} \;.
\label{eq:A0_def}
\end{equation}
The part of the second line of \Eqref{eq:BrignoleTrilinears}
involving $\tilde{K} \partial_{\bar\phi} \tilde{K}$ reproduces the
second line of \Eqref{eq:tri_coups_nonpr}.
Finally, the second term of \Eqref{eq:BrignoleTrilinears} contains 
$\partial_{\braket{\phi_m}} Y$ and yields the contribution 
$m_{3/2} \, T_{\tilde f^c_i \tilde F_j H_f} Y'_{f^c_i F_j H_f}$
to the trilinear coupling $a'_{\tilde f^c_i \tilde F_j H_f}$.  This term
can give a non-trivial family dependence even with a canonical K\"ahler
potential due to the non-trivial dependence of the Yukawa couplings on
the flavon fields.
Furthermore, this term does not depend on the unknown parameters
$\tilde\epsilon_f$ in the K\"ahler potential and hence it could be
directly linked to observable quantities.  However, the predictivity for
the physical trilinears is limited by two effects.

Firstly, the second line in \Eqref{eq:tri_coups_nonpr} depends on the
unknown parameter $\tilde\epsilon_f$.  This line also contains different
elements of the Yukawa matrices, which can be much larger than the
element appearing in the first line,
$Y'_{f^c_k F_j H_f} \gg Y'_{f^c_i F_j H_f}$ for some values of $k$.
Then Case~1 for $a'$ differs considerably from Case~2, since some terms
of the second line dominate for sufficiently large $\tilde\epsilon_f^2$.
For example, consider the element $a'_{\tilde f^c_2 \tilde F_3 H_f}$.
Up to Planck-suppressed terms, the second line of
\Eqref{eq:tri_coups_nonpr} reads
\[
- m_{3/2} c_2 \lambda_2^2\left[ \frac{\braket{\bar\phi_2}_2\braket{\bar\phi^\dagger_2}_2}{M_{\chi^f_2}^2} Y'_{f^c_2 F_3 H_f} + \frac{\braket{\bar\phi_2}_2\braket{\bar\phi^\dagger_2}_3}{M_{\chi^f_2}^2} Y'_{f^c_3 F_3 H_f}  \right]
\sim
- m_{3/2} c_2 \left[ \tilde\epsilon_f^2 \epsilon_f^2 - \tilde\epsilon_f^2 \right] .
\]
In Case 2 the parameter $\tilde\epsilon_f$ can easily be much larger
than $\epsilon_f$, so that the second line dominates over the first one,
which is proportional to the Yukawa coupling
$Y'_{f^c_2 F_3 H_f} \sim \epsilon_f^2$.  Thus, the order of magnitude of 
$a'_{\tilde f^c_2 \tilde F_3 H_f}$ is changed compared to Case 1.

Secondly, the transformation \eqref{eq:transftril} to the canonical basis
is controlled by the K\"ahler potential and thus again by the unknown
$\tilde\epsilon_f$.  In order to estimate the possible change, we choose
$V_{f^c}^\dagger$ as a lower-diagonal matrix
\cite{King:2004tx}.  Then the canonically normalised trilinears are
given by
\begin{equation}
    \hat a_{\tilde f_i^c \tilde F_j H_f} =
    (V_{f^c}^\dagger)_{ii}^{} \, a'_{\tilde f_i^c \tilde F_j H_f} +
    \sum_{k<i} (V_{f^c}^\dagger)_{ik}^{} \, a'_{\tilde f_k^c \tilde F_j H_f} \;,
\end{equation}
up to an overall (family-blind) factor (remember that here
$\tilde K_{F^\dagger F} \propto \mathbbm{1}$ and hence
$V_F \propto \mathbbm{1}$).
We have $(V_{f^c})_{ii} \sim 1$, 
$(V_{f^c})_{i\neq k} \lesssim \tilde\epsilon_f^2$, and
$a'_{\tilde f_k^c \tilde F_j H_f} \lesssim a'_{\tilde f_i^c \tilde F_j H_f}$
for $k<i$.
Thus, the corrections to each element of $a'$ are at most of the same
order of magnitude as the element itself as long as
$\tilde\epsilon_f<1$.  In other words, they only cause a change by an
$\mathcal{O}(1)$ factor
but do not change the power of $\epsilon_f$ appearing in the element.
Consequently, the effect of canonical normalisation is not terribly
different in the two limiting cases introduced above.
In conclusion, mainly due to the unknown size of
the second line of \Eqref{eq:tri_coups_nonpr} one loses predictivity for
the trilinear scalar couplings as well.

As a notable exception, the 33 element of the K\"ahler metric of the
right-handed matter fields is given by
\begin{equation} \label{eq:K33}
    \tilde K_{f^c_3 f^{c\dagger}_3} =
    1 + \xi_{f^c} +
    \lambda_3^2 \, (1+\xi_{\chi_2^f}) \,
    \frac{|\braket{\bar\phi_3}_3|^2}{M_{\chi^f_3}^2} +
    \lambda_3^4 \,
    \frac{|\braket{\bar\phi_3}_3|^4}{M_{\chi^f_0}^2 M_{\chi^f_3}^2} \;.
\end{equation}
As the large 33 entries of the Yukawa matrices are
\begin{equation} \label{eq:Y33}
    Y'_{f_3^c F_3 H_f} =
    \mathcal{N} \lambda_H \lambda_3^2 \,
    \frac{(\braket{\bar\phi_3}_3)^2}{M_{\chi_0^f} M_{\chi_3^f}} \sim 1
    \;,
\end{equation}
it follows that
$M_{\chi_0^f} \sim M_{\chi_3^f} \sim \braket{\bar\phi_3}$,
if we require the messengers $\chi_0^f$ and $\chi_3^f$ to be heavier
than $\braket{\bar\phi_3}$.%
\footnote{In the absence of a model for the messenger masses, even this is
 not predicted.  Depending on the structure of higher-order corrections
 to the effective K\"ahler potential, it is possible that
 $M_{\chi_0^f} M_{\chi_3^f} < \braket{\bar\phi_3}^2$ is sufficient
 for the consistency of the theory.
\label{fn:Loophole}}
Furthermore, the last term in \Eqref{eq:K33},
whose analogue was omitted in \Eqref{eq:Keff} because it is of
higher order in the messenger masses and thus not important for other
elements of $\tilde K_{f^c f^{c\dagger}}$, equals 
$|Y'_{f_3^c F_3 H_f} / \mathcal{N} \lambda_H|^2$.
Thus, with the qualification mentioned in footnote \ref{fn:Loophole}, one can indeed
gain some knowledge about the order of magnitude of the corrections to
the family-universal part $(1+\xi_{f^c})$ of
$\tilde K_{f^c_3 f^{c\dagger}_3}$.  
However, this is not the case for the remaining elements.  The smallness
of the Yukawa couplings other than $Y_{f_3^c F_3 H_f}$ implies a
hierarchy between the relevant flavon vevs and messenger masses, which
allows to vary these parameters significantly.

If the singlet messenger masses happen to be small enough to lead to
measurable deviations from the CMSSM due to large $\tilde\epsilon_f$, there could 
remain some predictions in the form of correlations between different
observables since the number of free parameters (including
$\tilde\epsilon_f$) is smaller than the number of independent soft SUSY
breaking parameters.  However, given the large number of parameters and
the complicated relations between model parameters and observables, it
is questionable if such predictions could in practice be found and
confirmed.
Another prediction that can arise in unified models is that the
parameters in the lepton sector are related to those in the quark
sector \cite{Baek:2000sj,Ciuchini:2003rg}.  However, this is a
consequence of the enlarged gauge symmetry rather than the family
symmetry.

\section{Improving Predictivity with Singlet Messengers}
\label{imppre}
\begin{figure}
\centering
\includegraphics{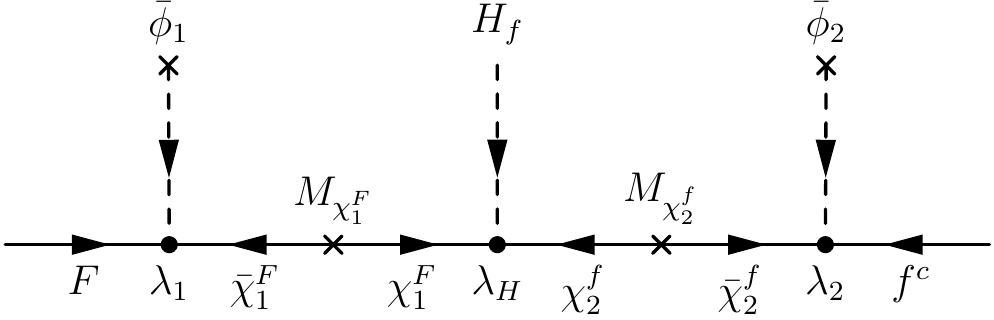}
\caption{Another possibility for generating Yukawa couplings.  All
 messengers are SU(3) singlets in this case.
}
\label{fig:Messengers23}
\end{figure}

One possible way to realise more predictions for the SUSY breaking parameters
is an extension of the theory that allows to restrict the
messenger masses, as proposed in \cite{King:2006me} for an SO(3) family
symmetry, or fixes the ratios between the flavon vevs.
Another way, which we will explore here, is to generate the
Yukawa couplings via diagrams
of the type shown in \Figref{fig:Messengers23}.%
\footnote{This kind of diagrams
 cannot be used for generating the 33 entries of the Yukawa couplings due
 to the quantum numbers assigned in \cite{deMedeirosVarzielas:2005ax}.
 Hence, we have to continue utilising a diagram of the type shown in
 \Figref{fig:Messengers33} to generate these particular entries.
}
In replacement of the messengers $\chi_0^f$, which are triplets under the
family symmetry SU(3) and singlets under $\SU(2)_\text{L}$, we now
employ the fields $\chi_1^F$ and $\chi_2^F$, which are SU(3) singlets
and $\SU(2)_\text{L}$ doublets. 

\begin{figure}
\centering
\includegraphics{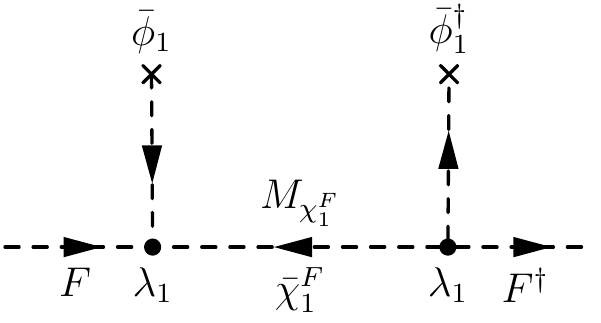} \;
\includegraphics{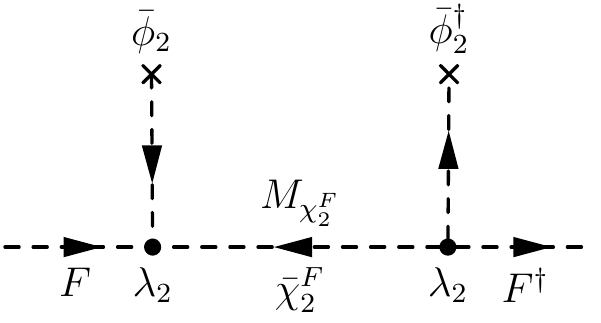}
\caption{Diagrams relevant for the K\"ahler potential for the alternative
 messenger sector of \Figref{fig:Messengers23}.
}
\label{fig:Kaehler23}
\end{figure}

We obtain the same
effective superpotential and Yukawa couplings as before, except that
$\chi^f_0$ is exchanged by $\chi^F_1$ and $\chi^F_2$, for instance
\begin{equation} \label{eq:Y23Pred}
    Y'_{f^c_2 F_3 H_f} =
    \mathcal{N} \lambda_H \lambda_2^2 \,
    \frac{\braket{\bar\phi_2}_2\braket{\bar\phi_2}_3}{M_{\chi^F_2} M_{\chi^f_2}}
    \equiv -\epsilon_f^2 \;,
\end{equation}
so that the phenomenology of the fermion sector is completely unchanged.
The diagram from \Figref{fig:Kaehler33} again yields contributions to
the K\"ahler potential of the right-handed matter fields
like in \Eqref{eq:EpsilonTilde}.
However, $\chi_n^F$ being family singlets, there arise the new diagrams
shown in \Figref{fig:Kaehler23}, which lead to non-universal corrections
to the K\"ahler potential of the left-handed fields as well, for example
\begin{equation}
    \tilde K_{F^\dagger_2 F_3} =
    \lambda_2^2 \, (1+\xi_{\chi_2^F}) \,
    \frac{\braket{\bar\phi_2}_2\braket{\bar\phi_2^\dagger}_3}{M_{\chi^F_2}^2} 
    \equiv -\tilde\epsilon_F^2 \;.
\end{equation}
Thus, \emph{all} messenger masses show up in the effective K\"ahler
potential and the soft parameters in this case.  
This means that although we still have considerable freedom to
adjust the expansion parameters $\tilde\epsilon_f$ and
$\tilde\epsilon_F$, we predict the correlation
\begin{equation} \label{eq:epsilonrelat}
    |\tilde\epsilon_F \, \tilde\epsilon_f| =
    \frac{(1+\xi_{\chi_2^F})^{1/2}\,(1+\xi_{\chi_2^f})^{1/2}}{\mathcal{N}\lambda_H}
    \, |\epsilon_f|^2 \;.
\end{equation}
Explicitly,
\begin{eqnarray} \label{eq:epsilonrelats}
    |\tilde\epsilon_Q \, \tilde\epsilon_u| \sim |\epsilon_u|^2
	\quad,\quad
    |\tilde\epsilon_Q \, \tilde\epsilon_d| \sim |\epsilon_d|^2
	\quad,\quad
    |\tilde\epsilon_L \, \tilde\epsilon_e| \sim |\epsilon_d|^2 \;,
\end{eqnarray}
where the appearance of the quark sector parameter $\epsilon_d$ rather than an
independent $\epsilon_e$ in the last relation is a particularity of the
model \cite{deMedeirosVarzielas:2005ax}.
As a consequence, no expansion parameter can be arbitrarily small, and
the breaking of the family symmetry produces off-diagonal elements in
all soft mass matrices.  For example,
we find using $\epsilon_d=0.13$ and $\tilde\epsilon_d<1$
\begin{equation}
    \tilde\epsilon_Q =
    \frac{\epsilon_d^2}{\lambda_H' \tilde\epsilon_d}
    \gtrsim 0.02 \;.
\end{equation}

The prediction \eqref{eq:epsilonrelat} still depends on the unknown
quantities $\lambda_H$, $\mathcal{N}$, and $\xi_\chi$.  By construction,
$\lambda_H \sim 1$,
like the other dimensionless couplings in the superpotential, so that
this parameter does not introduce a large uncertainty.  
The other unknown parameters are related to the hidden-sector field $h$.
Although they could be significantly larger than $1$ in principle, this
would require $\braket{h}\gg\Mp$ or a rather large number of additional
hidden-sector fields with vevs close to $\Mp$.  Thus, it seems
reasonable to expect these quantities to be of order unity as well.
For instance, for the Polonyi model \cite{Polonyi}
$\braket{h} = (\sqrt{3}-1) \, \Mp$, so that
$\mathcal{N} = \exp(\braket{h}^2/\,2\Mp^2) \approx 1.3$.

Some additional sources of uncertainty have not been included in the
above relations.
As mentioned earlier, we have used only three different couplings
$\lambda_1, \lambda_2$ and $\lambda_H$, and moreover neglected their
possible dependence on $h$.
Besides, we have defined $\epsilon_f$ before canonical normalisation.
This is not entirely correct, since these parameters are determined by a
fit to the \emph{physical}, canonically normalised Yukawa couplings.
However, canonical normalisation may not cause a significant change of
the Yukawa matrices, if the model is to predict the fermion masses.
Thus, as long as this condition is satisfied, canonical normalisation
does not produce more than another $\mathcal{O}(1)$ factor.

The soft SUSY-breaking parameters are computed as in the previous
section.  The soft masses depend on the parameters $\tilde\epsilon_f$
and $\tilde\epsilon_F$ but not on $\epsilon_f$, although of course now
they are related.
The trilinear couplings now include two terms that are not proportional
to the corresponding Yukawa couplings, the second line of
\Eqref{eq:tri_coups_nonpr} and in addition
\begin{equation}
a^\prime_{\tilde f^c_i \tilde F_j H_f} \supset - m_{3/2} c_1 \lambda_1^2 \frac{\langle \bar\phi_1 \rangle_{j} }{M^2_{\chi^F_1}} \sum_k \langle \bar\phi_1^\dagger \rangle_k \left(1-{\mathcal{O}}\left(\tfrac{|\braket{h}|^2}{\Mp^2}\right) 
\right) Y'_{f^c_i F_k H_f} \;,
\end{equation}
because of the off-diagonal term in the K\"ahler metric of the
left-handed matter fields. 

Leaving the particular model under consideration for a moment, an
obvious question is whether our results can be generalised to a simple
criterion for the predictivity of family models for the soft SUSY
breaking parameters.  We have seen that one can expect predictions, if
the masses of all messengers appear in the effective K\"ahler potential
at the order $M_\chi^{-2}$.  This is the case if a coupling
$F \bar\phi \chi$ or $f^c \bar\phi \chi$ exists for all messengers.
For $F,f^c \sim \rep{3}$ and $\bar\phi \sim \overline{\rep{3}}$, this
requires $\chi$ to be singlets (leaving aside representations with
dimension larger than $3$), as in the example of this section.  One
could also have $\bar\phi \sim \rep{3}$, though, which would allow a
coupling with triplet messengers.  Hence, one cannot conclude that
predictivity requires singlet messengers in general.
The converse statement evidently holds provided that 
$F\bar\phi,f^c\bar\phi \sim \rep{1}$: if all messengers are
singlets, then their masses appear in the effective K\"ahler potential.
However, non-Abelian family symmetries are often extended by extra
symmetries like $\U(1) \times \U(1)'$ in
\cite{deMedeirosVarzielas:2005ax} under which all messengers are
charged.  In other words, total singlets rarely exist, limiting the use
of this criterion.  Consequently, one usually cannot sidestep checking
the viability of each matter-flavon-messenger vertex.

\section{Comparison with Experimental Constraints} \label{sec:Exp}

In order to get an idea about the observable signatures that can be
expected, let us make a very rough estimate of the parameters relevant
for FCNC processes.  We assume a framework where the Yukawa couplings
are generated from diagrams of the type shown in
\Figref{fig:Messengers23}, with the exception of the $33$-entries.  For
the latter we employ the diagram of \Figref{fig:Messengers33}.  Thus,
the $33$-entries of the right-handed soft mass matrices and trilinear
couplings receive $\mathcal{O}(1)$ corrections.

 From \eqref{eq:CanonicalNormSoftMasses} it is straightforward to see that the same parameters $\tilde\epsilon_{Q}$ (in case of more than one flavon) parameterise both $V_F$ and $m^{' 2}_{\tilde F^\dagger \tilde F}$ and analogously the flavons of the type $\tilde\epsilon_{f}$ for $V_{f^c}$ and $m^{' 2}_{\tilde f^\dagger \tilde f}$ so the final expressions in \Eqref{eq:EstimateSoftMasses} are functions only of the original parameters and consequently flavour violation bounds constrain them. Although only indirectly, since before comparing to the appropriate bounds we need to go the so called SCKM (super CKM) basis where the Yukawa couplings are diagonal. So let us describe then the changes in the structure of these matrices.
One finds at
energies of the order $\braket{\bar\phi}$:
\begin{equation}
    \hat m^2_{\tilde f^\dagger \tilde f} \sim m^2_0 \begin{pmatrix}
    1 & \tilde\epsilon_f^2 \, \epsilon_d^2 &
     \tilde\epsilon_f^2 \, \epsilon_d^2 \\
    \cdot & 1 + \tilde\epsilon_f^2 & \tilde\epsilon_f^2 \\
    \cdot & \cdot & 1
    \end{pmatrix}
    \quad,\quad
    f = u,\, d,\, Q,\, e,\, L \;,
\end{equation}
where we have omitted the $\mathcal{O}(1)$ coefficients and where dots
stand for elements given by hermiticity.
For simplicity, we have
assumed a common prefactor $m_0^2$ in front of the matrix for all types
of sfermions.  Likewise, we will assume a unified gaugino mass
$m_{1/2}$,
so that in the limit $\braket{\bar\phi}\to0$ we have a special case of
the CMSSM\@. 

In elements $12$ and $13$ of the matrix above there appears the term
$\lambda_1^2 \braket{|\bar\phi_1|^2} / M^2_{\chi^f_1}$
(from our K\"ahler metric given in Eq.~\ref{explicitK}). Then for the correct structure of the Yukawa couplings (used in \cite{deMedeirosVarzielas:2005ax} and also needed for our Yukawa matrices) the relations $\braket{\bar\phi_1} \sim \epsilon_d \braket{\bar\phi_2}$ and $M_{\chi_1^f} \sim M_{\chi_2^f}$ are needed.

This form of  scalar masses was studied previously in
\cite{Antusch:2007re} for the case
$\tilde\epsilon_f=\epsilon_f$ for $f=u,d,e$ and
$\tilde\epsilon_Q=\tilde\epsilon_L=0$, i.e.\ diagonal mass matrices for
the $\SU(2)$ doublet sfermions.

We use the one-loop renormalisation group evolution estimate \cite{Martin:1997ns}
for the diagonal entries at low energy and neglect the running of the
off-diagonal entries. 
Up to now we have not mentioned the neutrino sector. The inclusion of it with the use of the see-saw mechanism has been widely used in $SU(3)$ models to reproduce the right spectra and mixing for oscillating neutrinos. In  most part of such models, the neutrino Yukawa couplings of $\mathcal{O}(1)$ do not influence the running because 
the corresponding right-handed singlet neutrino has a mass above $M_\text{GUT}$. 
 Later, we will briefly comment on scenarios where $Y_\nu$ plays a significant role.

Due to the hierarchical Yukawa couplings, the transformation to the SCKM basis is given to a very good approximation at low energy as follows:
\begin{equation}
    \widetilde m^2_{\tilde d,\text{RR}} \sim
    m_0^2 \begin{pmatrix}
     {\mathcal{R}}_{\tilde d,\text{RR}} & \tilde\epsilon_d^2 \, \epsilon_d &
      \tilde\epsilon_d^2 \, \epsilon_d + \epsilon_d^3 \\
     \cdot &  {\mathcal{R}}_{\tilde d,\text{RR}} & \tilde\epsilon_d^2 + \epsilon_d^2 \\
     \cdot & \cdot &  {\mathcal{R}}_{\tilde d,\text{RR}}
    \end{pmatrix} ,
\label{eq:m2dRR}
\end{equation}
where the factor $\mathcal{R}_{\tilde d,\text{RR}}$ corresponds to the RGE evolution increase at low energy. The matrices $\widetilde m^2_{\tilde Q,\text{LL}}$, $\widetilde m^2_{\tilde L,\text{LL}}$ and $\widetilde m^2_{\tilde e,\text{RR}}$ are analogous to \Eqref{eq:m2dRR}  with the replacements $\tilde\epsilon_d \to \tilde\epsilon_Q$,  $\tilde\epsilon_d \to \tilde\epsilon_L$ and $\tilde\epsilon_d \to \tilde\epsilon_e$ respectively and for the leptonic cases also the RGE factors ${\mathcal{R}}_{\tilde L}$ and ${\mathcal{R}}_{\tilde e}$ are different.

We present an example using the benchmark point SPS 1a, with values $m_0=100\GeV$, $m_{1/2}=250\GeV$, $A_0 = -100\GeV$ and $\tan\beta=10$, for which the estimate of \cite{Martin:1997ns} yields 
\begin{equation}
    (\tilde m_{\tilde q, LL}^2)_{ii} \sim 30 \, m_0^2
    \quad,\quad
    (\tilde m_{\tilde e, LL}^2)_{ii} \sim 4 \, m_0^2
    \quad,\quad
    (\tilde m_{\tilde e, RR }^2)_{ii} \sim 2 \, m_0^2 \;,
\end{equation}
i.e. ${\mathcal{R}}_{\tilde d,\text{RR}}\sim 30$, $\mathcal{R}_{\tilde e, LL} \sim 4$ and ${\mathcal{R}}_{\tilde e, RR} \sim 2$ for the quarks, lepton doublet and charged lepton singlets respectively.

Here we have ignored signs and assumed no severe cancellations, which can occur in fine-tuned cases.
We also neglect all complex phases, so that there are no contributions
to electric dipole moments and CP-violating parameters in meson mixing.

The quantities $\widetilde m^2_{\tilde u,\text{LL}}$ are less interesting due to the weaker experimental constraints (coming from $D$ rather than $K$ mixing).

We use the experimental constraints from 
$\Delta m_K$, $b \to s \gamma$, $\mu \to e \gamma$ etc.\ given in \cite{Ciuchini:2007ha, Altmannshofer:2009ne}.
In the mass insertion approximation, they can be translated into
constraints on the $\delta$ flavour violating parameters:
\bea
        (\delta^f_\text{RR})_{ij} &:=&
        \frac{(\widetilde m^2_{\tilde f,\text{RR}})_{ij}}
         {(\widetilde m^2_{\tilde f,\text{RR}})_{ii}},\nonumber \\
\left(\delta^f_{LR,RL}\right)_{ij} &: =&\frac{(\tilde m^2_{\tilde f,\text{LR,RL}})_{ij}}{\sqrt{(\tilde m^2_{\tilde f, \text{LL}})_{ii} (\tilde m^2_{\tilde f,\text{RR}})_{jj}}}.
\eea

\begin{table}
{\centering
\begin{tabular}{|ccc|}
\hline
&  { Our example} & { Bound} \\
\hline
\hline \vphantom{$\frac{{\epsilon_d^2}^2}{25_2}$}
        $(\delta^d_\text{RR})_{12}$ &
        $\frac{\widetilde\epsilon_d^{\,2}\,\epsilon_d}{30} \sim 7 \cdot 10^{-5}$ &
        $9 \cdot 10^{-3}$
\\ \hline \vphantom{$\frac{{\epsilon_d^2}^2}{25_2}$}
        $(\delta^d_\text{LL})_{12}$ &
        $\frac{\widetilde\epsilon_Q^{\,2}\,\epsilon_d}{30} \sim 7 \cdot 10^{-5}$ &
        $1 \cdot 10^{-2}$
\\ \hline
        $(\delta^d_\text{LR,RL})_{12}$
&
        $\frac{v}{\sqrt{1+\tan{^2 }\beta}}\left[-\frac{n A_0 \epsilon_d^3}{30 m_0^2}\right]\sim 4 n  \cdot 10^{-6}$ &
        $1 \cdot 10^{-5}$
\\ \hline \vphantom{$\frac{{\epsilon_d^2}^2}{25_2}$}
        $(\delta^d_\text{LL})_{23}$ &
        $\frac{\widetilde\epsilon_Q^{\,2}}{30} \sim 6 \cdot 10^{-4}$ &
        $2 \cdot 10^{-1}$
\\ \hline \vphantom{$\frac{{\epsilon_d^2}^2}{25_2}$}
        $(\delta^e_\text{LL})_{12}$ &
        $\frac{\widetilde\epsilon_L^{\,2}\,\epsilon_d}{4} \sim 6\cdot 10^{-4}$ &
        $6 \cdot 10^{-4}$
\\ \hline
        $(\delta^d_\text{LR,RL})_{23}$
&
        $\frac{v}{\sqrt{1+\tan{^2 }\beta}}\left[-\frac{n A_0 \epsilon_d^2}{30 m_0^2}\right] \sim 4 n \cdot 10^{-4}$ &
        $1 \cdot 10^{-3}$
\\ \hline
\end{tabular}
\\}
\caption{An example for the flavour violating parameters $\delta$ for the SPS 1a point, together with the corresponding experimental limit. For a detailed description of the formulas see the text in this section.}
\label{tab:Deltas}
\end{table}
Then the $(\delta^f_\text{XY})$ parameters are
given by
\bea
 \left(\delta^d_{LR,RL}\right)_{ij} &=& \frac{v}{\sqrt{1+\tan^2 \beta}}\left[-\frac{  \widetilde a_{\tilde d^c_i \tilde Q_j H_f}}{30 m_0^2}  +\frac{\mu^* \tan\beta \ \tilde{Y}^{\text{diag} }_{d^c_i Q_j}}{30 m_0^2\sqrt{2}} \right],\nonumber\\
\left(\delta^u_{LR,RL}\right)_{ij} &=& \frac{v\tan\beta}{\sqrt{1+\tan^2 \beta}}\left[-\frac{\widetilde a_{\tilde u^c_i \tilde Q_j H_f}}{30 m_0^2}  +\frac{\mu^*  \tilde{Y}^{\text{diag} }_{u^c_i Q_j}}{30 m_0^2 \tan\beta\sqrt{2}} \right],
\eea
where $\widetilde ~$ denotes the quantities in the SCKM basis.
Following our discussion on the transformation of the trilinear couplings $\hat a_{\tilde f^c_i \tilde F_j H_f}$ due to canonical normalisation, in our example the leading terms of the trilinear couplings are always proportional to the corresponding Yukawa coupling  $\hat Y_{\tilde f^c_i \tilde F_j H_f}$, i.e.\ the first line of \Eqref{eq:tri_coups_nonpr} dominates.
Then the SCKM transformation will not change the corresponding order of magnitude in $\hat a_{\tilde f^c_i \tilde F_j H_f}$:
\bea
  \widetilde a_{\tilde f^c_i \tilde F_j H_f} &=& \left[{U_\text{R}^f}^\dagger \hat a_{\tilde f^c \tilde F H_f} U_\text{L}^f\right]_{ij}=\left[{U_\text{R}^f}^\dagger 
    \tilde K_{H_f^\dagger H_f}^{-\frac{1}{2}}
    V_{f^c}^\dagger \, a'_{\tilde f^c \tilde F H_f} \, V_F U_\text{L}^f\right]_{ij}\nonumber\\
&=& {\mathcal{O}}\left(\left[T_{\tilde f^c_i \tilde F_j H_f}+T_h\frac{|\braket{h}|^2}{\Mp^2}\right] \hat Y_{\tilde f^c_i \tilde F_j H_f}\right) m_{3/2}.
\eea
Note that since we do not have 
the relations of the CMSSM case, 
we need to redefine an $A_0$ depending on each element of $a^\prime_{f^c_i f_j H_f}$ but, since $T_{\tilde f^c_i \tilde F_j H_f}$ are expected to be ${\mathcal{O}}(1)$, its order of magnitude can be estimated. What we have assumed in the numerical estimates in Table \ref{tab:Deltas}  is that we can express $(T_h\frac{|\braket{h}|^2}{\Mp^2}+T_{\tilde f^c_i \tilde F_j H_f})=n A_0 /m_{3/2}$ for a factor $n$ of ${\mathcal{O}}(1)$ that depends on each element $(i,j)$. In our example under consideration, $n$ can be a factor of a few
(i.e.\ $1+p_n^{ij}$ in \Eqref{eq:def_par_trilinears}) which can be well within the range to be probed by the forthcoming experiments.

In order to estimate the size of FCNCs in our setup, let us consider a
very simple example {\footnote{We have taken the numerical values of $\epsilon_d$ and $\epsilon_u$ from the latest fit \cite{Calibbi:2010rf} of the kind of Yukawa matrices we are using.}}:
\bea
        \widetilde\epsilon_Q = \widetilde\epsilon_L = \epsilon_d \approx 0.13
        \quad \Rightarrow \quad
        \widetilde\epsilon_d = \widetilde\epsilon_e = \epsilon_d \approx 0.13 
        \quad , \quad
        \widetilde\epsilon_u = \frac{\epsilon_u^2}{\epsilon_d} \approx 0.012
        \;.
\label{eq:examp_relations}
\eea
An example for the flavour violating parameters $\delta$ using the relations above is listed in \Tabref{tab:Deltas} for the SPS 1a point, together with the corresponding experimental limits.
We see that the constraints in the squark sector are easily satisfied for flavour violating parameters of the form $(\delta^d_\text{XX})$ but for $(\delta^d_\text{XY})$ we have an important dependence on what values are chosen for $A_0$, $m_0$ and $\tan\beta$. If $A_0$ is comparatively larger than $m_0$ then $(\delta^d_\text{XY})$ could be easily above the limit for it. Also
for a large $\tan\beta$
this could be a problem.
Taking $\epsilon_d\sim 0.15$, the values of $(\delta^d_\text{LR})_{12,23}$ are comfortably within the limits for the point SPS1a, while, for $A_0=-1100$ GeV, $m_0=200$ GeV and $\tan\beta=10$, $(\delta^d_\text{12})$ is at the limit but $(\delta^d_\text{23})$ satisfies all bounds.

The flavour violating parameters $\delta^u$ can be calculated analogously to those of $\delta^d$. However, its corresponding experimental are not so stringent \cite{Gabbiani:1996hi,Misiak:1997ei,Besmer:2001cj} compared to the $\delta^d$ constraints, and they are satisfied as long as the bounds on $\delta^d$ are satisfied with a choice of $\epsilon_u<\epsilon_d$.
We do not show $\delta^e_\text{RR}$, since they are only weakly constrained, too. Some tension can be seen in the lepton sector with $\mu \to e \gamma$,
consistently with what was found in \cite{Antusch:2007re}.

If there was a right-handed neutrino whose Yukawa coupling influenced the running of $(\hat m_{\tilde L^\dagger \tilde L}^2)_{ij}$ we could estimate its effect as \cite{GenCit:NRrun}:
\[(\delta \hat m_{\tilde L^\dagger \tilde L}^2)_{ij}= \frac{-1}{8\pi^2}(3 m^2_0+ A^2_0)\left[Y^{\nu\dagger} \log\left(\frac{\Mg}{M_N}\right) Y^\nu \right]_{ij}.\]
The form of $Y^\nu$ is unfortunately very model-dependent since we do not know the experimental value of the 
absolute scale and the nature of the oscillating neutrinos. Nevertheless, let us use a form that has been widely used in $SU(3)$ models, mainly that all elements of 
$Y^\nu$, except $Y^\nu_{11}$ and $Y^\nu_{33}$ are of $\mathcal{O}(\epsilon^3_\nu)$, where $\epsilon_\nu\in (\epsilon_u, 0.4)$ \cite{Calibbi:2010rf}. $Y^\nu_{11}< 
\mathcal{O}(\epsilon^3_\nu)$  and $Y^\nu_{33} \leq \mathcal{O}(1)$. Thus, by comparing the form of $(\delta \hat m_{\tilde L^\dagger \tilde L}^2)_{ij}$ to that of $(\hat 
m_{\tilde L^\dagger \tilde L}^2)_{ij}$, we see that for the $(1,2)$ and $(2,3)$ elements, respectively, we would compare \bea
&&\tilde\epsilon_{Q}^2\epsilon_d \ \text{vs.}\ \frac{-3}{8\pi^2}\epsilon^6_\nu (l_2+l_3)\nonumber\\
&&\tilde\epsilon_{Q}^2\ \text{vs.}\ \frac{-3}{8\pi^2}\epsilon^3_\nu (l_3)\nonumber,
\eea
where $l_i=\log\left(\frac{\Mg}{M_{Ni}}\right)\sim {\mathcal{O}}(1)$. Hence only for $(\hat m_{\tilde L^\dagger \tilde L}^2)_{23}$ we could have a sizable contribution from the right-handed neutrinos when considering values of $\epsilon_\nu\approx 0.4$.\footnote{
The situation changes when $A_0$ is large. First of all, the simple one loop approximation to $(\delta m_{\tilde L^\dagger \tilde L}^2)_{ij}$ given above does not describe properly the running effects any more and indeed they could play a special role in enlarging or reducing this contribution \cite{Petcov:2003zb}, therefore they may play a more important role. However this needs to be analysed using the exact running whose detailed numerical study is left for our forthcoming work.}

\section{Conclusions \label{con}}

The potential of family symmetries to predict the soft supersymmetry
breaking parameters in addition to the fermion masses has been studied
actively in recent years
\cite{Ross:2002mr,Ross:2004qn,Antusch:2007re,Antusch:2008jf,Calibbi:2009ja,Calibbi:2010rf,Olive:2008vv}.
In this work, instead of using effective potentials restricted only by the symmetries of a model as has been conventionally done, we have explored an ultra-violet (UV) completion 
to study all the ingredients for the family symmetry breaking and its connection to supersymmetry breaking in supergravity.
The specification of the UV completion helped us in clarifying how the flavon and the messenger fields contribute to the soft supersymmetry breaking parameters and Yukawa couplings. 
We argued that, in a conventional model with triplet messenger fields,
predictions for the soft parameters are hindered 
because they depend on unknown parameters
that are not fixed by fitting the Yukawa couplings to experimental data.
Those parameters, for example the messenger masses, are associated with the non-canonical K\"ahler potential
and the hidden sector breaking supersymmetry.

As one possibility to improve the situation, we have proposed a model where all the messengers are family singlets. 
This allowed us to derive predictions in the form of correlations
between different soft parameters.
Such models with predictive power robust enough to test the 
underlying family symmetry would deserve further examination in view of the wealth of forthcoming experimental data probing flavour and CP violation.

\subsection*{Acknowledgements}
We would like to thank Steve King, Oleg Lebedev, Jan Louis, and Oscar Vives for helpful
discussions.  This work was supported by the German Science Foundation
(DFG) via the Junior Research Group ``SUSY
Phenomenology'' within the Collaborative Research Centre 676
``Particles, Strings and the Early Universe'', by the Michigan Center
for Theoretical Physics, and by the INFN\@.
We thank the MCTP, the Galileo Galilei
Institute for Theoretical Physics, the University of
Hamburg, and the Abdus Salam ICTP for their hospitality.

\appendix
\section{K\"ahler Metric}

With all the previous assumptions, to a good approximation the K\"ahler metric of the right-handed matter fields for the model of \Secref{sec:spec_example} is given by
\bea
\tilde K_{f^c f^{c\dagger}}=\mathbbm{1}+\left[
\begin{array}{ccc}
  \frac{\lambda^{f2}_1 \braket{|\bar\phi_1|^2}}{M^2_{\chi^f_1}}\quad & \frac{\lambda^{f2}_1 \braket{|\bar\phi_1|^2}}{M^2_{\chi^f_1}} & \frac{\lambda^{f2}_1 \braket{|\bar\phi_1|^2}}{M^2_{\chi^f_1}}\\
\frac{\lambda^{f2}_1 \braket{|\bar\phi_1|^2}}{M^2_{\chi^f_1}} \quad & \frac{\lambda^{f2}_2 \braket{|\bar\phi_2|^2}}{M^2_{\chi^f_2}} + \frac{\lambda^{f2}_1 \braket{|\bar\phi_1|^2}}{M^2_{\chi^f_1}} \ & \frac{\lambda^{f2}_2 \braket{|\bar\phi_2|^2}}{M^2_{\chi^f_2}} + \frac{\lambda^{f2}_1 \braket{|\bar\phi_1|^2}}{M^2_{\chi^f_1}}\\
\frac{\lambda^{f2}_1 \braket{|\bar\phi_1|^2}}{M^2_{\chi^f_1}} \quad & \frac{\lambda^{f2}_2 \braket{|\bar\phi_2|^2}}{M^2_{\chi^f_2}} +\frac{\lambda^{f2}_1 \braket{|\bar\phi_1|^2}}{M^2_{\chi^f_1}} \ & \frac{\lambda^{f2}_3 \braket{|\bar\phi_3|^2}}{M^2_{\chi_{3}^f}} + \frac{\lambda^{f2}_2 \braket{|\bar\phi_2|^2}}{M^2_{\chi^f_2}} + \frac{\lambda^{f2}_1 \braket{|\bar\phi_1|^2}}{M^2_{\chi^f_1}} 
\end{array}
 \right].
\label{explicitK}
\eea
$|\bar\phi_a|^2_{i,j}\equiv(\bar\phi_a)_i (\bar\phi_a)^\dagger_j$, are the $i$ and $j$ components of $\bar\phi_a$ and $\bar\phi_a^\dagger$ respectively.
We have assumed that 
\begin{equation} \label{eq:Hierarchy2}
\frac{(\braket{\bar\phi_3}_3)^2}{M^2_{\chi_3^f}} \gg
\frac{(\braket{\bar\phi_2}_3)^2}{M^2_{\chi_2^f}} \gg
\frac{(\braket{\bar\phi_1}_3)^2}{M^2_{\chi_1^f}},
\end{equation}
in addition to \Eqref{eq:condit_flav_hier}.

\frenchspacing
\providecommand{\bysame}{\leavevmode\hbox to3em{\hrulefill}\thinspace}

\end{document}